\documentclass[amsthm]{elsart}
\usepackage{amsmath,amsfonts,amssymb}
\usepackage{graphicx}
\usepackage[square]{natbib}
\journal{Physics Letters A}
\newtheorem{theorem}{Theorem}
\newtheorem{lemma}{Lemma}
\newtheorem{proposition}{Proposition}

\newtheorem{remark}{Remark}
\newcommand\mvector{\boldsymbol}

\newcommand\vd{\mvector{d}}

\newcommand\vp{\mvector{p}}
\newcommand\vq{\mvector{q}}

\newcommand\vx{\mvector{x}}

\newcommand\vA{\mvector{A}}

\newcommand\vE{\mvector{E}}
\newcommand\vF{\mvector{F}}
\newcommand\vG{\mvector{G}}

\newcommand\vK{\mvector{K}}

\newcommand\vX{\mvector{X}}

\newcommand\vzero{\mvector{0}}
\newcommand\valpha{\mvector{\alpha}}
\newcommand\vLambda{\mvector{\Lambda}}
\newcommand\vvarphi{\mvector{\varphi}}
\newcommand\vvareps{\mvector{\varepsilon}}

\newcommand\cG{{\mathcal G}} 
\newcommand\cH{{\mathcal H}}

\newcommand\cX{{\mathcal X}}

\newcommand\field{\mathbb}

\newcommand\R{\field{R}}

\newcommand\C{\field{C}}

\newcommand\N{\field{N}}
\newcommand\Q{\field{Q}}

\newcommand\bbP{\mathbb{P}}

\newcommand\CP{\ensuremath{\C\bbP}}

\newcommand\Dt{\frac{\mathrm{d}\phantom{t} }{\mathrm{d}\mspace{1mu} t}}

\begin{document}
\begin{frontmatter}
 \title{Finiteness of integrable $n$-dimensional homogeneous polynomial
potentials}
\thanks{This research has been supported by  the
European Community project GIFT  (NEST-Adventure Project no. 5006) and by Projet
de l'ANR N$^\circ$~JC05$_-$41465.}
\author{Maria Przybylska}
\ead{mprzyb@astri.uni.torun.pl}
\address{Institut Fourier, UMR 5582 du CNRS,
Universit\'e de Grenoble I,
100 Rue des Maths, BP 74, 38402 Saint-Martin d'H\`eres Cedex, France}
\address{Toru\'n Centre for Astronomy,
  N.~Copernicus University, 
  Gagarina 11, PL-87--100 Toru\'n, Poland.}

\begin{abstract}
 We consider natural Hamiltonian systems of $n>1$ degrees of freedom 
 with 
polynomial homogeneous potentials of degree $k$. We show that under a
genericity assumption, for a fixed $k$, 
 at most only a finite number of such systems is integrable. We also 
explain how to 
find explicit forms of these integrable potentials for small $k$.
\end{abstract}

\end{frontmatter}

\section{Introduction}
\label{sec:intro}

At least  half of the models which appear in physics, astronomy and
 other applied sciences have a form of a system of ordinary 
differential equations depending usually on several parameters. 
The question if the considered system possesses one or more first 
integrals is fundamental. First integrals give conservation laws 
for the model. Moreover, from an operational point of view, they 
simplify investigations of the system. In fact, we can always lower 
the dimension of the system by the number of its independent first 
integrals.  If we know a sufficient number of first integrals, 
we can solve explicitly the considered system. As a rule, except 
possible obvious first integrals, as  Hamiltonians for  Hamilton's  
equations, additional first integrals exist only for specific values 
of parameters of the considered systems. Thus, the problem is how to 
find these values of parameters, or, how to show that the system does 
not admit any additional first integral for specific values of 
the parameters. The problems mentioned above are generally very hard, 
and  in spite of their basic physical importance there are no
universal methods to solve them  even for  very special 
classes of differential equations.

In past the search for first integrals was based on the direct 
method due to  Darboux, see e.g. \cite{Whittaker:65::}. Applying this
 method, we postulate a general form of the first integral. Usually, 
this first integral  depends on some unknown functions. The condition 
that it is constant along   solutions of the analysed system gives 
rise to a set of partial differential equations determining the 
unknown functions.  Complexity of  the obtained  partial 
differential equations is  the reason why  it is usually assumed 
that the first integral is a polynomial with respect to momenta of 
low degree. For more information about the direct method 
see~\cite{Hietarinta:87::}.

In the sixties of the previous century, Ablowitz, Ramani and Segur
  \cite{Ablowitz:80::b,Ablowitz:80::a} proposed a completely different
 method of searching for integrable systems.  The  kernel of this method, 
originating from the  works of Kovalevskaya
 \cite{Kowalevski:1888::,Kowalevski:1890::} and Painlev\'e
 \cite{Painleve:1902::}, is a conjecture that solutions of 
integrable systems  after  the extension to the complex time plane 
should be still simple, more precisely, single-valued. If all solutions
 of a given  system are single-valued, then we say that  it possesses 
the Painlev\'e property. But, to check if  a given system possesses the
 Painlev\'e property, we must have at our disposal a single-valued 
particular solution of the system (or its appropriate truncation). Then
 the necessary condition for the Painlev\'e property is following: 
all solutions of the variational equations along this single-valued particular
solution 
are single-valued. If for some specific values of parameters the 
considered system has the Painlev\'e property, then, assuming those 
values of parameters, we can look for first integrals applying the 
direct method.  This means that the Painlev\'e property has played the 
role of necessary integrability conditions and, for this reason,  it is 
sometimes called the Painlev\'e test. The results of Kovalevskaya and 
Lapunov \cite{Kowalevski:1888::,Kowalevski:1890::,Lyapunov:1894::} 
showed that checking the Painlev\'e property is in fact reduced to 
checking if a certain matrix, the so-called Kovalevskaya matrix 
(see the next section), is semisimple, and  its eigenvalues, the 
so-called Kovalevskaya exponents, are integers. Yoshida 
 \cite{Yoshida:83::a} showed  that Kovalevskaya exponents are related 
to the  degrees of first integrals,  and this fact simplifies the 
second step of the analysis, namely, finding the explicit form of  the 
fist integral.   The Painlev\'e test appeared to be very effective and 
 many new integrable systems were found thanks to its application. The 
main advantage of this method  is its simplicity.  Its weak point  is 
the fact that there is no rigorous proof that the Painlev\'e property 
is directly related to the integrability. In fact, there are known 
examples of  integrable systems that do not pass the Painlev\'e test, 
by this reason the weak Painlev\'e  test was introduced, see 
\cite{Ramani:82::,Ramani:89::,Grammaticos:97::}.

Let us remark that in Hamiltonian mechanics there exist a few other 
tools for testing the integrability,  see \cite{Kozlov:96::}, however, 
they usually work  for very restricted classes of Hamiltonian systems.

Quite recently two mathematically rigorous approaches  to the 
integrability problem formulated by 
Ziglin \cite{Ziglin:82::b,Ziglin:83::b} and Morales-Ruiz and 
Ramis \cite{Morales:01::a,Morales:99::c} have appeared.  They explain  
relations between the existence of first integrals and branching of 
solutions as functions of the complex time and give  necessary 
integrability conditions  for Hamiltonian systems. It appears that the 
integrability  is related to   properties of the monodromy group or 
the differential Galois group of variational equations along a 
particular solution.

In this paper we apply the Morales-Ramis approach to the  Hamiltonian 
systems defined in a linear symplectic space, e.g.,
$\R^{2n}$ or $\C^{2n}$ equipped with  canonical variables 
$\vq=(q_1,\ldots,q_n)$,
$\vp=(p_1,\ldots,p_n)$, and given by a natural Hamiltonian function 
\begin{equation}
H=\dfrac{1}{2}\sum_{i=1}^np_i^2+V(\vq).
\label{eq:ham}
\end{equation}
We assume that $V(\vq)$ is a homogeneous polynomial of degree $k>2$. 
 The integrability of Hamiltonian systems with Hamiltonian 
\eqref{eq:ham}
was analysed by the direct method, the Painlev\'e analysis and some 
other techniques, see
\cite{Hietarinta:83::,Hietarinta:87::,Ramani:89::,Lakshmanan:93::}. 
Nevertheless, a quick overview of the literature shows that except for 
some
``easy''
cases only sporadic examples of integrable systems with two or three 
degrees of
freedom governed by the Hamiltonian of the form~\eqref{eq:ham} were found. 
In all
integrable cases first integrals are polynomials and their degrees 
with respect to
the momenta are not greather than four. Hence, it is natural to ask: 
do we know
all integrable systems with Hamiltonian~\eqref{eq:ham}? It is hard to 
believe
that
the answer to this question is positive. In fact, as far as we know, 
in all works
only very limited families of such systems were investigated. Thus,
what can we expect? Are there infinitely many integrable Hamiltonian 
systems
which
wait to be discovered? 

The aim of this note is to give a necessarily limited answer to the above
question. The main result of this paper shows that assuming that 
potential $V$ is generic, the number
of meromorphically integrable systems with Hamiltonian
  ~\eqref{eq:ham} is finite.  

Let us explain here what does it mean a generic
potential. 
Hamilton's equations generated by~\eqref{eq:ham} admit particular solutions of
the form
\begin{equation}
 \vq(t) = \varphi(t)\vd, \qquad \vp(t) =\dot \varphi(t)\vd,  
\end{equation}  
provided $\vd$ is a nonzero solution of
\begin{equation}
\label{eq:vdp}
 V'(\vd)=\vd,
\end{equation} 
and $\varphi(t)$ satisfies $\ddot \varphi = -\varphi^{k-1}$.   A direction
$\vd\in\C^n$ defined by a solution of~\eqref{eq:vdp} is called a  Darboux point
of potential $V$. We say that potential $V$ is generic iff it admits exactly $k$
different  Darboux points. For details see Section~\ref{sec:DPKMR}. 

To prove our finiteness result we combine the Morales-Ramis theory and  a kind of global
 Kovalevskaya analysis  of the auxiliary system
\begin{equation}
 \label{eq:aux1}
 \Dt \vq = V'(\vq).
\end{equation}
It appears that the Kovalevskaya exponents of the above system are 
closely related to the integrability of Hamiltonian system given
 by~\eqref{eq:ham}. The Morales-Ramis theory gives strong restrictions 
on their values. On the other hand, we can calculate the Kovalevskaya 
exponents for different particular solutions of~\eqref{eq:aux}. The key
 point is the fact that the Kovalevskaya exponents calculated for 
different solutions are not arbitrary, i.e., there exist certain 
relations among them.

Just to avoid missunderstanding let us fix terminology here.  We 
consider complex
Hamiltonian systems with phase space $\C^{2n}$ equipped with the 
standard
canonical structure. First integrals are always assumed to be 
meromorphic in
appropriate domains. By saying that a potential $V$ is integrable, 
we understand
that
the Hamilton equations  generated by Hamiltonian~\eqref{eq:ham} are 
integrable
in the Liouville sense. It is easy to check that if potential
 $V(\vq)$ is
integrable, then also $V_{\vA}(\vq):=V(\vA\vq)$ is integrable for an 
arbitrary
$\vA\in\mathrm{GL}(n,\C)$ satisfying $\vA\vA^T=\alpha \vE$, $\alpha\in\C^\star$
and $\vE$ is the identity matrix, see e.g. \cite{Hietarinta:87::}.
 Potentials $V$ and $V_{\vA}$ are called equivalent, and the set of 
all potentials is
divided into disjoint classes of equivalent potentials.   Later a 
potential
means a class of equivalent potentials in the above sense.

The plan of this paper is following. In the next section we briefly 
recall basic facts from the Kovalevskaya analysis and the Morales-Ramis
 theory. In Section~3 we formulate and prove our main results. In the 
last section we explain how our approach can be used for a systematic 
analysis of the integrability of homogeneous potentials.
 
\section{Darboux points, Kovalevskaya exponents and Morales-Ramis theory}
\label{sec:DPKMR}
At the beginning  we
remind basic notions of the Kovalevskaya-Painlev\'e analysis, for more 
details and
references see \cite{Kozlov:96::}.

Let us consider a polynomial system 
\begin{equation}
 \label{eq:hds}
 \Dt \vx = \vF(\vx), \qquad \vx \in\C^n,
\end{equation} 
with homogeneous right hand sides $\vF=(F_1,\ldots,F_n)$, $\deg F_i=k$, $k>1$,
for $i=1,\ldots, n$. A direction $\vd\in\C^n$ (i.e., a non-zero vector)
 is
called a Darboux point of system~\eqref{eq:hds} if  $\vd$ is  parallel to
$\vF(\vd)$ and $\vF(\vd)\neq\vzero$. Note that a Darboux point can be 
considered
as a point $[d_1:\cdots:d_n]$ in projective space $\CP^{n-1}$.  The set of all
Darboux points for system~\eqref{eq:hds}  is denoted by $\mathcal{D}_{\vF}$.  It
can be empty, finite or infinite. If all Darboux points are isolated, then
$\mathcal{D}_{\vF}$ is finite, and in  a generic case  it 
has $D(n,k):=(k^n-1)/(k-1)$ elements, see Proposition~4 on page 348
in~\cite{Guillot:04::}.
We always normalise Darboux points of system~\eqref{eq:hds} in such a way
that they satisfy  the following
nonlinear equations
\begin{equation}
 \label{eq:dp} \vF(\vd)=\vd,
\end{equation} 
but we must remember that different solutions of the above equations 
can define
the same Darboux
point. 
\begin{remark}
\label{rem:ndar}
 Let us notice that if $\vd\neq\vzero$ is a solution of equation~\eqref{eq:dp},
then by homogeneity of $\vF$, also $\widetilde \vd:= \varepsilon \vd$ is a solution of this equation provided
$\varepsilon$ is a $(k-1)$-th root of the unity. Thus if equation~\eqref{eq:dp}
has $m$ different solutions, then they define only  $m/(k-1)$  different Darboux
points.
\end{remark}

The Kovalevskaya matrix $\vK(\vd)$ at a Darboux point $\vd\in
\mathcal{D}_{\vF}$ is defined as 
\begin{equation}
 \label{eq:Kd}
 \vK(\vd):=\vF'(\vd)-\vE,
\end{equation} 
where $\vF'(\vd)$ is the Jacobian matrix of $\vF$ calculated at $\vd$. 
Eigenvalues $\Lambda_i=\Lambda_i(\vd)$, $i=1,\ldots,n$,
 of the Kovalevskaya matrix   $\vK(\vd)$ are called the Kovalevskaya 
exponents.
Using the homogeneity of $\vF$ it is easy to prove that one of the 
Kovalevskaya
exponents, let us say $\Lambda_n$, is $k-1$. We call this
eigenvalue trivial.

If the general solution of system~\eqref{eq:hds} is single-valued, 
then
the Kovalevskaya exponents should be integer. However, as we have 
already mentioned the Painlev\'e test is not a correct integrability 
condition. The first strict relation between the existence of a first 
integral and the Kovalevskaya exponents was found by Yoshida
\cite{Yoshida:83::a}, who 
proved that if system  \eqref{eq:hds} possesses a polynomial first 
integral whose gradient does not vanish at the Darboux point, then the 
degree of the homogeneity of this first integral belongs to the 
spectrum of the Kovalevskaya matrix. This result was later generalised 
in \cite{Furta:96::,Nowicki:96::,Goriely:96::} and the final relation 
is the following. If system~\eqref{eq:hds}
possesses a polynomial or rational first integral, then the Kovalevskaya
exponents calculated at a certain Darboux point satisfy a resonance 
relation. However, in this paper we do not
use  this connection between the Kovalevskaya exponents and the  
integrability in the  class of polynomial or rational functions 
but we use a  stronger result concerning the integrability in the 
wider class of meromorphic functions.

As we said, the idea of Kovalevskaya  gave a strong impulse for 
searching a
relation
between the  integrability and branching of solutions as
functions of the complex time. This, somewhat mysterious, relation was 
fully
explained for Hamiltonian systems by an elegant and powerfull theory of
S.~L.~Ziglin \cite{Ziglin:82::b,Ziglin:83::b}. The basic idea of this 
theory is
following.  To have a chance to describe possible branching, we need a
particular single-valued solution $\vvarphi(t)$ of the considered system. 
Knowledge about solutions close to $\vvarphi(t)$ comes from the 
variational
equations along $\vvarphi(t)$. The monodromy group of these equations 
describes
branching of solutions  close to $\vvarphi(t)$.  The existence of 
integrals of
the system puts a restriction on the monodromy group---it cannot be too
``big''. 
At the end of the previous century, the strength of the Ziglin theory was
considerably improved thanks to the application of the differential Galois
theory \cite{Put:03::}. The Morales-Ramis theory, see \cite{Morales:99::c}, is a
kind of an algebraic version of the  Ziglin theory---instead of the monodromy
group the differential Galois group of the variational equations 
is used to find
obstructions for the integrability. The main theorem of the 
Morales-Ramis theory
states that if the investigated system is integrable
in the Liouville sense, then the identity component of the
differential Galois group of the variational equations along a
particular solution is Abelian, see \cite{Morales:01::a,Morales:99::c}.

Morales-Ruiz and Ramis used their theory to give the  strongest known 
necessary conditions for the
integrability of Hamiltonian systems with the homogeneous 
potential ~\eqref{eq:ham}.  Let us describe
them shortly.  To apply the Morales-Ramis theory we
need a particular solution. As we explained in Introduction assuming that the
considered potential has a Darboux point, such a solution is given
by~\eqref{eq:vdp}.
It gives a family of phase curves $\Gamma_\varepsilon$ of the form 
\begin{equation*}
 \dot\varphi^2=\frac{2}{k}(\varepsilon -\varphi^k), \qquad \varepsilon\neq 0
\end{equation*}
These curves are elliptic for $k=3,4$ and hyperelliptic for $k>4$.

It is easy to show
that the variational equations along solution~\eqref{eq:vdp} have the form
$\ddot \vx=
-\varphi(t)^{k-2}V''(\vd)\vx$. Thus, assuming that the Hessian matrix $V''(\vd)$
is diagonalisable, we can find coordinates $(y_1, \ldots, y_n)$  such that in
these variables 
equations read $\ddot
y_i=  -\lambda_i\varphi(t)^{k-2}y_i$, for $i=1,\ldots,n$, where 
$\lambda_1,
\ldots,\lambda_n$ are eigenvalues of $V''(\vd)$.  As was observed by 
Yoshida
\cite{Yoshida:87::a}, the following change of the independent variable 
$t\rightarrow z:=\varphi(t)^k/\varepsilon$ transforms $i$-th 
variational equation into the
Gauss hypergeometric equation with parameters dependent on
 $k$ and $\lambda_i$. 
But, for the hypergeometric equation the monodromy, as well as the 
differential Galois
groups, are well known, see e.g., \cite{Iwasaki:91::}. 
Basing on these facts, 
J.~J.~Morales-Ruiz and J.~P.~Ramis formulated in \cite{Morales:01::a} a
general theorem
concerning the integrability of Hamiltonian systems with a homogeneous
potential.  Here, we formulate this theorem for a polynomial
homogeneous potential.
\begin{theorem}
\label{thm:MoRa}
If the Hamiltonian system given by~\eqref{eq:ham} with the polynomial 
homogeneous
potential $V(\vq)$ of degree $k>2$ is meromorphically integrable in
the Liouville sense, then  values of
$(k,\lambda_i)$ for $i=1,\ldots,n$ belong to the following list
\begin{center}
\begin{tabular}{clcl}
1.& $\left( k, \dfrac{k}{2}p(p-1)+p\right)$, 
&2.& $\left(k,\dfrac {k-1} {2k}+p(p+1)\dfrac{k}{2}\right)$, \\[1em]
3.& $\left(3, \dfrac 1 {6}\left( 1 +3p\right)^2-\dfrac 1 {24}\right)$, &
4.& $\left(3,\dfrac 3 {32}\left(  1  +4p\right)^2-\dfrac 1 {24}\right)$,
\\[1em]
5.& $\left(3,\dfrac 3 {50}\left(  1  +5p\right)^2-\dfrac 1 {24}\right)$,&
6.& $\left(3,\dfrac{3}{50}\left(2 +5p\right)^2-\dfrac 1 {24}\right)$,\\[1em]
7.& $ \left(4,\dfrac{2}{9} \left( 1+ 3p\right)^2-\dfrac 1 8 \right)$,&
8.& $\left(5,\dfrac 5 {18}\left(1+ 3p\right)^2-\dfrac 9 {40}\right)$,\\[1em]
9.& $\left(5,\dfrac 1 {10}\left(2+5p\right)^2-\dfrac 9 {40}\right)$,&
 & 
\end{tabular}
\end{center}
where $p$ is an integer. 
\end{theorem}
Let us notice that one eigenvalue of $V''(\vd)$,  let us say
$\lambda_n$, is  $k-1$, so it does not give any restriction to the
integrability. For a typical situation when the investigated potential 
depends
on some parameters, using the above theorem, we are able to 
distinguish infinite
families (depending on parameters)  of potentials which are suspected to be
integrable.
\begin{remark}
If $V'(\vd)=\vd$,  then $\widetilde\vd = \gamma \vd$ satisfies $V'(
\widetilde\vd)=\gamma^{k-2}\widetilde\vd$, and using $\widetilde\vd$ we can find
a particular solution as we did with $\vd$. Although eigenvalues of
 $V''(\vd)$
and $V''(\widetilde\vd)$ are different, we do not obtain a new 
restriction for
the
integrability. The reason of this is the fact that $\vd$ and $ \widetilde\vd$
define the same phase curves. 
\end{remark}
The above remark justifies  introducing  the notion of a Darboux point 
of a
homogeneous polynomial potential $V$. We say that $\vd\in\C^n$ is a Darboux
point of a 
 $V\in\C[\vq]$, if it is a Darboux point of the auxiliary
system
\begin{equation}
 \label{eq:aux}
 \Dt \vq = V'(\vq).
\end{equation}
We always normalise coordinates of a Darboux point $\vd$
in
such way that they satisfy~\eqref{eq:vdp}.  We denote by $\mathcal{D}_V$ the set
of
Darboux points of system~\eqref{eq:aux}. Notice that 
for $\vd\in \mathcal{D}_V$,
the Kovalevskaya exponents $\Lambda_i(\vd)$ are given by
 $\Lambda_i(\vd) =
\lambda_i(\vd)-1$, where $\lambda_i(\vd)$ are eigenvalues of $V''(\vd)$. 

\section{Main result}
\label{sec:MR}
It is obvious that the more Darboux points we have for a given 
potential, the
more obstructions for its integrability follow from 
Theorem~\ref{thm:MoRa}.  
Investigating the integrability of two dimensional potentials, see
\cite{Maciejewski:04::g,Maciejewski:05::b}, we noticed the following 
fact. For a
generic case the number of Darboux points is $k=\deg V$. 
We have two eigenvalues of $V''(\vd_i)$ for every $\vd_i\in\mathcal{D}_V$.  One
of
them
is $k-1$, the remaining one we denote by $\lambda_i$. It appears that 
non-trivial eigenvalues $\lambda_1,\ldots, \lambda_k$ calculated at all Darboux
points  are not 
independent, i.e., they satisfy a certain relation.
This relation has the simplest form if we express it in terms of
$\Lambda_i=\lambda_i-1$, i.e., in terms of the Kovalevskaya exponents 
of the
auxiliary system~\eqref{eq:aux}, and  it  reads
\begin{equation}
\label{eq:relk}
\sum_{i=1}^k \frac{1}{\Lambda_i} = -1.
\end{equation}
Now, if the system is integrable, then $\Lambda_i=\lambda_i-1$ take 
values determined by
Theorem~\ref{thm:MoRa}. In \cite{Maciejewski:05::b} we showed that for 
an
arbitrary $k>2$ there exist at most a finite number of $\Lambda_1, \ldots,
\Lambda_k$, satisfying this requirement. Moreover, from our 
considerations in
\cite{Maciejewski:05::b} it follows that   the number
of potentials of degree $k$ which have specified values of $(\Lambda_1, \ldots,
\Lambda_k)$, is  finite. All the above facts imply that for a fixed
$k>2$, the number of  integrable homogeneous potentials of degree $k$ 
with the
maximal number of Darboux points is finite.  Thus, it is natural to 
ask if we
can prove a similar fact for a case when $n>2$. Unfortunately, the 
methods we
used
in \cite{Maciejewski:05::b} are applicable only when $n=2$.  Now our aim  is to
show how to overcome this difficulty. 

A theorem proved in
\cite{Guillot:04::} plays the central role in our considerations. Here 
we formulate it in a form
adapted to our needs. 
Let us return to a general first order homogeneous 
system~\eqref{eq:hds} and
assume that it has the maximal number of Darboux points. As it was 
mentioned,
one
of the Kovalevskaya exponents at an arbitrary Darboux point $\vd$ is 
$k-1$;
we denote the remaining ones by $\vLambda(\vd)=(\Lambda_1(\vd),\ldots,
\Lambda_{n-1}(\vd))$.  Let  $\tau_i$ for $0\leq i \leq n-1$, denote
the elementary symmetric polynomial in $(n-1)$ variables  of degree $i$ i.e.
\[
\tau_r(\vx):=\tau_r(x_1,\ldots,x_{n-1})=\sum_{1\leq i_1<\cdots<i_r\leq
n-1}\prod_{s=1}^r x_{i_s}, \qquad 1\leq r\leq n-1. 
\]
and $\tau_0(\vx):=1$.
The theorem below is in fact a reformulation of Corollary~12 on page 359
in~\cite{Guillot:04::}.
\begin{theorem}
\label{thm:kojot}
Assume that system \eqref{eq:hds} with homogeneous polynomial right 
hand sides
of degree $k$ has the maximal number of Darboux points
 and let $S$ be a symmetric homogeneous polynomial  in
$n-1$ variables of degree 
less than  $n$. Then, the number   
\begin{equation}
\label{eq:Rkojot}
R:
=\sum_{\vd\in\mathcal{D}_{\vF}}\frac{S(\vLambda(\vd))}{\tau_{n-1}(\vLambda(\vd))
},
\end{equation}
depends only on the choice of $S$, dimension $n$ and homogeneity
degree $k$.
\end{theorem}
In other words, $R=R(S,n,k)$ does not depend on a specific choice of
 $\vF$,
provided that $\vF$ has the maximal number of Darboux points. The above
 theorem
shows that there exist $n-1$ different universal relations among
 ``non-trivial''
Kovalevskaya exponents calculated at all Darboux points. 
In order to use the above theorem effectively, we have to know the 
values of
$R(S,n,k)$ for arbitrary  $n$, $k$ and a chosen set of $n$
independent symmetric homogeneous polynomials $S_i$ of degree $ i$ for
$i=0,\ldots, n-1$. In \cite{Guillot:04::} one can find these values 
for $n=3$
and $k=2$.
 Fortunately, the method used in \cite{Guillot:04::} works also
 in the general
case. To calculate $R(S,n,k)$  it is enough to choose a system for 
which  one
can easily determine 
the Kovalevskaya exponents, but the system must
be defined for arbitrary $n>2$ and $k>2$, and, of course, it must have 
the
maximal number of Darboux points. These requirements are satisfied by 
the 
$n$-dimensional generalisation of the Jouanolou system
\begin{equation}
 \label{eq:ju}
 \dot x_i = x_{i+1}^k,  \qquad 1\leq i\leq n, \quad x_{n+1}\equiv x_i,
\end{equation}
see \cite{Maciejewski:00::b}.  For this system the Kovalevskaya exponents do not
depend on a Darboux point and can be written explicitly. We show this in the
following lemma. 
\begin{lemma}
Let $\vLambda=(\Lambda_1,\ldots,\Lambda_{n-1})$ denotes the non-trivial
Kovalevskaya exponents  calculated at a Darboux point of system~\eqref{eq:ju}.
Then the elementary symmetric polynomials of $\vLambda$ take
the following values
\begin{equation}
 \label{eq:taur}
 \tau_r(\vLambda)=(-1)^r \sum_{i=0}^{r}
\binom{n-i-1}{r-i}k^{i},
\end{equation} 
for $ 0\leq r \leq n-1$. 
\end{lemma}
\begin{proof}
Solutions $\vd=(d_1,\ldots,d_n)$ of equation \eqref{eq:dp} describing Darboux
points can be written as
\begin{equation}
d_n=s,\qquad  d_{n-1}=s^k,\qquad d_{n-2}=s^{k^2},\ldots,d_2=s^{k^{n-2}},\qquad
d_1=s^{k^{n-1}},
\label{eq:dabcie}
\end{equation}
where $s$ is a  primitive root of unity of degree $k^n-1$, i.e. $s$ is a
solution of the cyclotomic equation
\begin{equation}
 s^{k^n-1}-1=0.
\label{eq:dub}
\end{equation}
Equation \eqref{eq:dabcie} has $k^n-1$ complex solutions, hence by
Remark~\ref{rem:ndar} system~\eqref{eq:ju} has
 $(k^n-1)/(k-1)$  Darboux points. 
The Kovalevskaya matrix  at point $\vd$ has the form
\[
 \vK(\vd)=\begin{pmatrix}
      -1&kd_2^{k-1}&0&\cdots&\cdots&0\\
      0&-1&kd_3^{k-1}&0&\cdots&0\\
\vdots&\ddots&\ddots&\ddots&\ddots&\vdots\\
\vdots&\cdots&\ddots&-1&\ddots&0\\
0&\cdots&\cdots&0&-1&kd_n^{k-1}\\
kd_1^{k-1}&0&\cdots&\cdots&0&-1
     \end{pmatrix}.
\]
Thus its characteristic polynomial is 
\begin{equation}
\begin{split}
 &P(\Lambda)=\operatorname{det}\,(\vK(\vd)-\Lambda\vE)=(-1)^n(\Lambda+1)^n+(-1)^
{n-1}
k^nd_1^{k-1}\cdots d_n^{k-1}\\
&=(-1)^n[(\Lambda+1)^n-k^n].
\end{split}
\label{eq:charac}
\end{equation}
In this way we showed that all Darboux points have the same Kovalevskaya
exponents
given by
\begin{equation}
 \Lambda_i = k \varepsilon^{n-i} -1, \qquad  1\leq i\leq n,
\end{equation}  
where $\varepsilon$ is a primitive $n$-th root of the unity. 
In order to find elementary symmetric functions of nontrivial eigenvalues
$\vLambda:=(\Lambda_1,\ldots,\Lambda_{n-1})$ we   factorise characteristic  
polynomial
\eqref{eq:charac} in the following way
\[
P(\Lambda)=(-1)^n(\Lambda+1-k)Q(\Lambda,n,k),
\]
where
\[
 Q(\Lambda,n,k)=\sum_{p=0}^{n-1}\,\sum_{i=p}^{n-1}
\binom{i}{p}k^{n-1-i}\Lambda^p=\sum_{q=0}^{n-1}\,\,\,\sum_{i=n-1-q}^{n-1}
\binom{i}{n-1-q}k^{n-1-i}\Lambda^{n-1-q}.
\]
Thus symmetric functions $\tau_r(\vLambda)$ are up to  the sign  coefficients of
the above polynomial 
\[
\tau_r(\vLambda)=(-1)^r \sum_{i=n-1-r}^{n-1}
\binom{i}{n-1-r}k^{n-1-i},\qquad r=0,\ldots, n-1. 
\]
From the above formula we obtain~\eqref{eq:taur} by a simple change of indices. 
\end{proof}

By the above lemma we have in particular  $\tau_1(\vLambda) =1-n-k$,
 and 
\begin{equation}
\tau_{n-1}(\vLambda)=(-1)^{n-1}\frac{k^n-1}{k-1},\quad \tau_{n-2}(\vLambda)
=(-1)^{n}\frac{k^n-n(k-1)-1}{(k-1)^2}.
\end{equation} 
Knowing \eqref{eq:taur} we can calculate $R(S,n,k)$ for an arbitrary choice of
$S$. In what follows we need explicit formulas for $S  =\tau_1^r$  and 
$S=\tau_r$ for $0\leq r\leq n-1$.  
\begin{proposition}
For  $ 0\leq r\leq n-1$ we have
\begin{equation}
\label{eq:Rt1r}
 R(\tau_1^r,n,k) = (-1)^{n-1}(1-n-k)^r, 
\end{equation}
and
\begin{equation}
 R(\tau_{r}, n, k )=(-1)^{r +1-n}\sum_{i=0}^{r}
\binom{n-i-1}{r-i}k^{i}.
\end{equation} 
\end{proposition}
\begin{proof}
 It is enough to insert formulas~\eqref{eq:taur} into~\eqref{eq:Rkojot} and
perform elementary simplification. 
\end{proof}
Let us notice that in particular we have 
\begin{equation}
 R(\tau_{n-2},n,k)=-\frac{k^n-n(k-1)-1}{(k-1)^2}.
\end{equation}

Now, the question is if we can use the above facts for our problem. 
As it was
mentioned, the integrability conditions of a homogeneous potential
 are given in
terms of the Kovalevskaya exponents of the auxiliary gradient
system~\eqref{eq:aux}.   The $\C$-linear space of polynomial 
homogeneous systems
of a given degree has the dimension greater than the space of 
polynomial
homogeneous
gradient systems of the same degree. Thus it seems to be possible that 
the number of
isolated Darboux points of system~\eqref{eq:aux} with the potential of degree
$k$ is always smaller than $D(n,k-1)$. 
But it is not like that---there exist potentials of degree $k$ which have
$D(n,k-1)$ Darboux points.
The simplest example is following
\begin{equation}
\label{eq:V}
 V_0 =\sum_{i=1}^n q_i^k.
\end{equation} 
We prove  that potentials of degree $k$ with $D(n,k-1)$ Darboux
points are generic. Let us precise the meaning of a generic potential. 
The $\C$-linear space $\cH_k$ of all homogeneous polynomials of degree $k$  has
dimension 
\begin{equation*}
 d = \binom{n+k-1}{k}.
\end{equation*}
Let us fix an monomial ordering $\prec$ of variables $q_1,\ldots, q_n$. Then
every homogeneous polynomial $V$ of degree $k$ can be uniquely written in the
form 
\begin{equation*}
 V=\sum_{i=1}^dv_i \vq^{\valpha_i},
\end{equation*}
where $\vq^{\valpha_1}\prec \cdots\prec \vq^{\valpha_d}$ are all monomials of
degree $k$. Hence we identify $\cH_k$ with $\C^d$ identifying $V$ with $(v_1,
\ldots, v_d)\in\C^d$. We convert $\cH_k$ into a complete normed space fixing  in
$\C^d$  an arbitrary norm. 
Now we show the following.
\begin{lemma}
 \label{lem:gen}
Let $\cG_k\subset \cH_k$ be a set of all homogeneous potentials of degree $k>2$
such that 
\begin{enumerate}
 \item if $V\in\cG_k$, then $V$ has maximal number of Darboux points $\vd_1,
\ldots \vd_s$, where 
\begin{equation*}
 s = D(n,k-1):=\frac{(k-1)^n-1}{k-2},
\end{equation*}
and
\item for each Darboux point all the Kovalevskaya exponents are different from
zero.
\end{enumerate}
Then set $\cG_k$ is open and non-empty.
\end{lemma}
\begin{proof}
First we show that $\cH_k$ is not empty. It is an easy exercise to check that
$V_0$ satisfies condition (1) and (2) so it is an element of $\cG_k$. 

To prove that that $\cG_k$ is open we have to show that for every $V\in\cG_k$
all potentials close enough to $V$ also belongs to $\cG_k$. To this end we
notice that a Darboux point $\vd$  of $V\in\cH_k$ is a zero of 
\begin{equation}
 \vG(\vq):=V'(\vq)-\vq.
\end{equation} 
We claim that if $V\in\cG_k$, then a Darboux point $\vd$ of $V$  is an isolated
zero of $\vG$. In fact,  the Jacobian of $\vG$ calculated at $\vd$ is not
singular as $\vG'(\vd)=\vK(\vd)$ and by assumption $\det \vK(\vd)\neq0$.   Let
$V\in\cG_k$ and 
\begin{equation*}
 W=\sum_{i=1}^d\varepsilon_i \vq^{\valpha_i}.
\end{equation*} 
Darboux points of $V+W$ are solutions $\vd(\vvareps)$ of 
\begin{equation*}
 \vG(\vd,\vvareps):=V'(\vd)+W'(\vd) -\vd=\vzero, \qquad
\vvareps=(\varepsilon_1,\ldots,\varepsilon_d).
\end{equation*}
By assumption that $V\in\cG_k$ we have that for $\vvareps=\vzero$ the above
equation has $s$ isolated solutions 
$\vd(\vzero)$.  Hence for $\Vert{\vvareps}\Vert$ small enough  there exist $s$
solutions $\vd(\vvareps)$. This exactly means that for an arbitrary $V\in\cG_k$
there exists an open subset of  $\cG_k$ containing $V$, i.e., $\cG_k$ is open.
\end{proof}
 Thanks to the above  lemma we can apply directly Theorem~\ref{thm:kojot} to the
auxiliary system and 
this gives the following.
\begin{theorem}
\label{thm:Vrel}
Let us assume  that a homogeneous polynomial potential $V\in \C[\vq]$ of degree
$k$ has  $D(n,k-1)$ Darboux points $\vd\in\mathcal{D}_V$. Then non-trivial
Kovalevskaya exponents $\vLambda(\vd)$ satisfy the following relations:
\begin{equation}
 \label{eq:rkoj}
 \sum_{\vd\in\mathcal{D}_V}\frac{ \tau_1(\vLambda(\vd))^r
 }{\tau_{n-1}(\vLambda(\vd))}=(-1)^{n-1}(2-n-k)^r,
\end{equation} 
 or, alternatively
\begin{equation}
 \label{eq:rksym}
 \sum_{\vd\in\mathcal{D}_V}\frac{ \tau_r(\vLambda(\vd))
 }{\tau_{n-1}(\vLambda(\vd))} =
(-1)^{p}\sum_{i=0}^{r}\binom{n-i-1}{r-i}(k-1)^{i},
\end{equation} 
 for $0\leq r\leq n-1$.
\end{theorem} 
Now, if a Hamiltonian system~\eqref{eq:ham} with potential
 $V$ satisfying assumptions of the above 
theorem is integrable, then   $\Lambda_i(\vd)$ for $1\leq i\leq n-1$ 
and $\vd\in
\mathcal{D}_V$ take rational values given in Theorem~\ref{thm:MoRa} and
 satisfy
relations~\eqref{eq:rkoj} and \eqref{eq:rksym}. These are really strong
restrictions. This fact is shown  by the following theorem.
\begin{theorem}
 \label{thm:my}
 Among Hamiltonian systems given by~\eqref{eq:ham}  with  homogeneous potentials
of fixed degree $k>2$ admitting the maximal number of Darboux
points only a finite number is integrable.
\end{theorem}
In order to prove this theorem we recall Lemma~B.1 from
\cite{Maciejewski:05::b}. 
\begin{lemma}
\label{lem:dum}
Let us consider the following equation 
\begin{equation}
\label{eq:eqX} 
X_1 + \cdots + X_m = c, \qquad c>0,
\end{equation} 
and look for its solutions $\vX=(X_1,\ldots, X_m)\in\cX^m$ where $\cX$
is  a set of all sequences $\{x_i\}_{i\in\N}$  of non-negative real numbers such
that 
  $\lim_{n\rightarrow \infty} x_n=0$.   
Then for an arbitrary $c>0$ equation~\eqref{eq:eqX} has at most a finite
number of solutions in $\cX^m$.
\end{lemma}
To prove Theorem~\ref{thm:my} we only need  
one relation given in Theorem~\ref{thm:Vrel}.
Namely, relation \eqref{eq:rksym} for $r=n-2$ reads
\begin{equation}
 \label{eq:invL}
 \sum_{\vd\in\mathcal{D}_{V}}\sum_{i=1}^{n-1}\frac{1}{\Lambda_i(\vd)}=
-\frac{(k-1)^n-n(k-2)-1}{(k-2)^2}.
\end{equation} 
Let us define  $(n-1)D(n,k-1)$ quantities $ X_i= 1/\Lambda_i(\vd)$ where
$i=1,\ldots,n-1$ and $\vd\in\mathcal{D}_{V}$.
Then from Theorem~\ref{thm:MoRa} 
it follows that for a fixed $k$,  $X_i$ belong to an appropriate  set $\cX_k$ possessing the properties

\begin{enumerate}
\item $\cX_k=\{ x^{(k)}_n \in\Q\setminus\{0\}\;|\;
  x^{(k)}_n\in(-\infty,-1)\cup(0,\infty), \quad n\in\N \} $,
\item For each $k$ the sequence $\{ x^{(k)}_n \}$ has only one
  accumulation point at 0.  In particular, for each $k$ only a finite
  number of $x^{(k)}_n$ take negative values not greater than -1.
\end{enumerate}
From relation~\eqref{eq:invL} it follows that at least one of $X_i$ is
negative. However, if $X_i$ is negative, then it cannot be  greater than -1.
Hence, not all of $X_i$ are negative. So assume that $X_i$ for $i=m+1,
\ldots, l$, where $l=(n-1)D(n,k-1)$, are negative for some $0<m<l$.  There is
only finitely many
choices for $X_{m+1}, \ldots, X_l$. For each of them we can rewrite
relation~\eqref{eq:invL} in the form \eqref{eq:eqX}. But, by the 
above lemma, this means that \eqref{eq:invL} has only a finite number of
solutions.

Let us note that for $n=2$ relation \eqref{eq:invL} transforms into
\eqref{eq:relk}.
\section{Discussion and Comments}
\label{sec:CO}
Fact that there exist some relations between the Kovalevskaya exponents was
observed  earlier  in a study of Painlev\'e property of multi-parameter systems,
see e.g. relation~(3.12) in~\cite{Grammaticos:83::}.  For two degrees of freedom
and a homogeneous potential of degree 3, one can find all Darboux points and the
respective Kovalevskaya exponents explicitly and then check directly that
relation~\eqref{eq:relk} holds.  Exactly in this way relation~\eqref{eq:relk}
was found for $k=3$ in~\cite{Maciejewski:04::g}. However this method fails for
$k>4$ as there is no way to find explicit solutions of non-linear
equations~\eqref{eq:vdp}. For an arbitrary $k$ relation~\eqref{eq:relk} was found
in~\cite{Maciejewski:05::b} but the method used there cannot be directly
generalised to  higher dimensional systems. Nevertheless it allows to find
certain relations beetween the Kovalevskaya exponents for non-generic
potentials.

Let us underline that Theorem~\ref{thm:my} is only one, 
and not the most important,
 consequence of relations \eqref{eq:rkoj} and \eqref{eq:rksym}.
These relations  give  also 
 a possibility to investigate completely the integrability problem of 
potentials with a given degree $k$ in an algorithmic way.   At first, for a given
$k$ we have to find all solutions 
$\Lambda_{i,j}=\Lambda_i(\vd_j)$, $i=1,\ldots,n-1$, $j=1,\ldots, D(n,k-1)$  of
relations \eqref{eq:rkoj} or \eqref{eq:rksym} such that the corresponding
quantities $\lambda_{i,j}= \Lambda_{i,j}+1$ belong to an item in the table given
in Theorem~\ref{thm:MoRa}. Such solutions are called admissible. For this a
computer algebra program is needed. For $n=2$ and relatively small values of $k$
such solutions can be find quickly. Tables~\ref{tab:lambdy3} and
\ref{tab:lambdy4}, taken from~\cite{Maciejewski:04::g,Maciejewski:05::b} give all
such solutions for $k=3$ and $k=4$, respectively. 
\begin{center}
\begin{table}[h]
\begin{tabular}{l}
\hline 
\hline
 $\{\Lambda_1,\Lambda_2, \Lambda_3\}$ \\
\hline 
\hline
$\{-1,-1,1\}$ \\ 
$\{-2/3,4,4\}$ \\ 
$\{-7/8,14,14\}$ \\
$\{-2/3,7/3,14\}$ \\
\hline
\end{tabular}
\\
\caption{Admissible solutions of \eqref{eq:relk} for $k=3$.}
\label{tab:lambdy3}
\end{table}
\end{center}
\begin{center}
\begin{table}[h]
\begin{tabular}{l}
\hline 
\hline
$\{\Lambda_1,\Lambda_2, \Lambda_3,\Lambda_4\}$ \\
\hline 
\hline
$\{-1,-1,2,2\}$ \\ 
$\{-5/8,5,5,5\}$ \\
$\{-5/8,2,20,20\}$ \\
$\{-5/8,27/8,27/8,135\}$\\
$\{-5/8,2,14,35\}$  \\
\hline
\end{tabular}
\\
\caption{Admissible  for solutions of \eqref{eq:relk}
$k=4$.}
\label{tab:lambdy4}
\end{table}
\end{center}
It must be said however that for bigger values of $k$ finding all admissible
solutions start to be a computer time demanding problem.  For $k>5$ it is known
that relation~\eqref{eq:relk} has always at least two solutions.
\begin{equation*}
 A_{k}^{(1)}=\{-1,-1,\underbrace{k-2,\ldots,k-2}_{k-2\
\text{times}}\},\qquad
A_{k}^{(2)}=\left\{-\frac{k+1}{2k},\underbrace{k+1,\ldots,k+1}_{k-1\
\text{times}}\right\}.
\end{equation*} 
However for $k=14, 17, 19, 26, \ldots$ additional solutions appear, see
\cite{mp:06::h}. For $n>2$ finding all admissible solutions of \eqref{eq:rkoj}
or \eqref{eq:rksym} is a very difficult problem even for $k=3$.

In the next step we have to find  the  potentials which give admissible
Kovalevskaya exponents.
 As we have already mentioned, for a given admissible solution $\{\Lambda_{ij}\}$ the number of the corresponding non-equivalent potentials is finite. A procedure of potentials reconstruction
reduces to finding all solutions of a system on polynomial equations. For $n=2$
and small $k$ this problem can be solved explicitly. For bigger values of  $k$
we  do not know if it is possible to find explicit solutions. 
For $n>2$ finding all admissible solutions of \eqref{eq:rkoj} or
\eqref{eq:rksym} is a very difficult problem even for $k=3$. The algorithm
applied for $n=2$ is useless and new   more efficient one  is needed.
For $A_{k}^{(1)}$ and $A_{k}^{(2)}$ one can find the corresponding potentials solving
certain linear differential equations, for details see~\cite{mp:06::h}.

Among selected potentials we can find integrable, as well 
as not integrable ones.
At this point we need  a tool stronger than Theorem~\ref{thm:MoRa} to 
prove that those non-integrable are really non-integrable. 
Fortunately, there exists such  a tool. It is  yet another theorem due to 
Morales and Ramis which says that if the system is integrable, then 
the identity component of the differential Galois group of the $i$-th 
order variational equations is Abelian for any $i\in\N$. For more details see
\cite{Morales:99::c,Morales:00::a}. However this theorem can be used effectively
only when $k=3$ or $k=4$.  For example of applications and details see~\cite{Maciejewski:04::g}.

We have already mentioned
that 
the integrability is a highly non-generic phenomenon. 
Our Theorem~\ref{thm:my} concerns only
potentials
 with the maximal number of Darboux points.  But of course there exist
potentials with 
not isolated Darboux points, potentials which have no   the maximal number of
isolated Darboux points, 
and  potentials without Darboux points.  Among them one can find integrable
ones. An example of an integrable potential without Darboux points is given  in
\cite{mp:05::d}. In this example the additional first integral  is of the fourth
degree with
 respect
to the momenta.  Nevertheless, we 
conjecture that Theorem~\ref{thm:my} is true without any assumptions.

In the end we mention that  relations \eqref{eq:rkoj} and \eqref{eq:rksym},
and their generalisations for nongeneric cases, can be derived in  a way 
different from 
 that used in~\cite{Guillot:04::}. It is exposed, together with an 
integrability 
 analysis of  three dimensional potentials, in our forthcoming 
paper~\cite{mp:06::i}. 

%

\newcommand{\noopsort}[1]{}\def\polhk#1{\setbox0=\hbox{#1}{\ooalign{\hidewidth
  \lower1.5ex\hbox{`}\hidewidth\crcr\unhbox0}}} \def\cprime{$'$}
  \def\cydot{\leavevmode\raise.4ex\hbox{.}} \def\cprime{$'$} \def\cprime{$'$}
  \def\cprime{$'$} \def\cprime{$'$} \def\cprime{$'$}
  \def\polhk#1{\setbox0=\hbox{#1}{\ooalign{\hidewidth
  \lower1.5ex\hbox{`}\hidewidth\crcr\unhbox0}}} \def\cprime{$'$}
  \def\cprime{$'$}

\end{document}